# Multi-Knob Switchable Chiral Superconductivity Quartet in Rhombohedral Graphene

Zhenqi Hua[1, †], Shenyong Ye[2, †], Phatthanon Pattanakanvijit[2, †], Gang Shi[1], Tonghang Han[2], Emily Aitken[2], Jixiang Yang[2], Junseok Seo[2], Haoyang Liu[1], Ran Hao[1], Kaitai Xiao[1], Jiaxing Guo[1], Võ Tiến Phong[1, 3], Kenji Watanabe[5], Takashi Taniguchi[6], Chunli Huang[7], Cyprian Lewandowski[1, 3, 4], Long Ju[2,*], Peng Xiong[1, 3, 4,*], and Zhengguang Lu[1, 3, 4,*]

1. *Department of Physics, Florida State University, Tallahassee, FL 32306*
2. *Department of Physics, Massachusetts Institute of Technology, Cambridge, MA 02139, USA*
3. *FSU Quantum Initiative, Florida State University, Tallahassee, FL 32306, USA*
4. *National High Magnetic Field Laboratory, Tallahassee, FL 32310*
5. *Research Center for Electronic and Optical Materials, National Institute for Materials Science, 1-1 Namiki, Tsukuba 305-0044, Japan*
6. *Research Center for Materials Nanoarchitectonics, National Institute for Materials Science, 1-1 Namiki, Tsukuba 305-0044, Japan*
7. *Department of Physics and Astronomy, University of Kentucky, Lexington, Kentucky 40506-0055, USA*

† Z.H., S.Y., and P.P., contributed equally to this work

*zl13b@fsu.edu, pxiong@fsu.edu, longju@mit.edu

**Chiral superconductors break orbital time-reversal symmetry and may host topological quasiparticles with non-Abelian statistics[1–3]. In rhombohedral graphene, superconductivity develops from a spin-valley-polarized quarter-metal (QM) parent state and features unique magnetic hysteresis of resistance that indicates orbital time-reversal-symmetry-breaking[4]. Exploring and controlling the full spin-valley flavors of such superconductivity could enable novel superconducting and topological devices[5–8], but have remained unexplored. Here we report transport measurements on rhombohedral hexalayer graphene (R6G), which reveal a new superconducting state (SCH) that is induced by an out-of-plane magnetic field, in addition to CSC similar to those observed in thinner layers. This SCH state emerges above 0.8 T, persists up to 1.6 T and can be switched on/off by magnetic field $\boldsymbol{H_{\perp}}$, carrier density $\boldsymbol{n}$, and gate displacement field $\boldsymbol{D}$. Quantum oscillations and anomalous Hall measurements show that SCH stems from a field-induced quarter-metal (QM′) parent phase, which carries orbital magnetization opposite to that of the zero-field QM. Across the full $(\boldsymbol{n}, \boldsymbol{D}, \boldsymbol{H_{\perp}})$ parameter space, superconductivity can be realized from all four spin-valley isospin flavors, establishing a switchable chiral-superconductor quartet in R6G. We interpret the parent-state switching as arising from competition between a Kane-Mele like spin-valley splitting[9] and magnetic-field coupling to spin-valley-dependent magnetic moments. Our work establishes rhombohedral graphene as a multi-knob platform for different isospin-polarized superconductivities, which enables programmable superconducting networks with possible Majorana modes along domain walls[10,11].**

Rhombohedral graphene has emerged as a highly tunable platform for correlated and topological electronic phases[4,12–14], largely because its low-energy bands carry multiple spin, valley and layer degrees of freedom that can be selected by interactions and external fields[15–17]. This rich isospin structure has already appeared across a wide range of phases of matter, including spin- and valley-polarized half- and quarter-metals[16,17], layer-antiferromagnetic and layer-polarized insulating states[18–21], charge-ordered phases[22,23], and gate-tunable Chern insulators[24–26]. In many of these cases, carrier density $n$, displacement field $D$ and magnetic field $H$ provide complementary knobs for selecting among competing isospin-polarized states[15–17,20,27,28]. The possibility of such rich flavors and their multi-knob control in superconducting states remain largely unexplored in rhombohedral graphene, but are highly desirable for exploring novel gate-defined superconducting junctions and topological superconducting domain walls.

Recent experiments in rhombohedral graphene have established chiral superconductivity (CSC), which is characterized by non-zero orbital angular momentum that spontaneously breaks time-reversal and mirror symmetries. In this work, we use the term CSC in a phenomenological sense to denote superconducting states with macroscopic orbital time-reversal-symmetry breaking, independent of the microscopic pairing mechanism or topological classification. These superconducting states exhibit hysteretic magnetic switching between time-reversal-related states near zero field, corresponding to quarter-metal (QM) parent states with opposite valley polarizations[4]. However, this binary magnetic switching does not yet exploit the full spin-valley flavors available in rhombohedral graphene, nor does it establish electrical switching of superconducting states with different isospin flavors. The rich isospin phenomenology already seen in the metallic and insulating phases therefore motivates the search for multi-flavor superconductivity, in which distinct superconducting states can be selected by $n$, $D$, $H$ within a single crystalline device.

Here we present transport measurements on dual-gated rhombohedral hexalayer graphene (R6G) devices (Fig. 1a and Extended Data Fig. 1). At near-zero out-of-plane magnetic field $H_\perp$, we observe two superconducting states, SC1 and SC2, whose critical temperatures and hysteretic magnetic response resemble the CSC reported in rhombohedral tetra- and pentalayer graphene[4]. Upon increasing $H_\perp$, SC2 is suppressed, while a new superconducting state we name as SCH emerges next to SC1 above $\mu_0 H_\perp^* \sim 0.8$ T. SC1 and SCH are separated by a finite-resistance boundary and can be switched by $n$, $D$, and $H_\perp$. Quantum oscillations and anomalous Hall measurements show that SCH develops from a field-induced quarter-metal parent state QM′, distinct from the low-field QM parent state of SC1. We attribute the QM–QM′ switching to competition between Kane–Mele-like spin-orbit coupling and $H_\perp$ coupling to spin-valley-dependent magnetic moments, which reorders the four isospin flavors and produces a field- and gate-selectable superconducting quartet in R6G.

## Zero-Field CSC in R6G

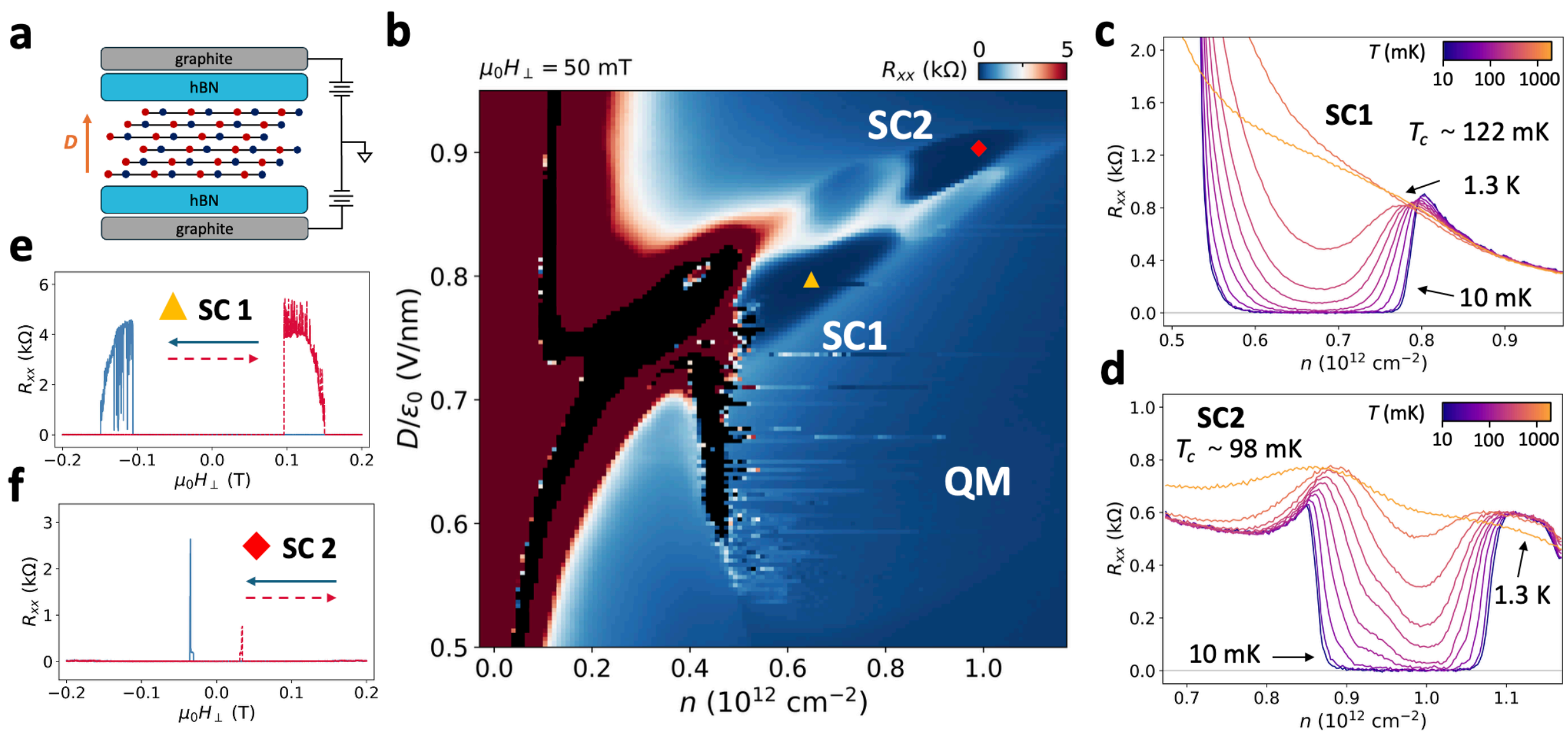

***Fig. 1 | Chiral superconductivity (CSC) in rhombohedral hexalayer graphene (R6G). a****, Schematic of the device structure, in which the R6G is intentionally misaligned by large twist angles with adjacent hBN layers to avoid the moiré superlattice effect.* ***b****, Four-terminal longitudinal resistance ($R_{xx}$) of R6G as a function of $n$ and $D$ taken at a small applied out-of-plane magnetic field of 50 mT to suppress valley fluctuations, and base temperature (10 mK at the mixing chamber). Two zero resistance regions, labeled as SC1 and SC2, are observed adjacent to a quarter-metal (QM) state.* ***c****,* ***d****, Temperature dependence $R_{xx}$ linecuts as function of $n$ through SC1 (****c****) and SC2 (****d****). The SC1 and SC2 traces are taken at $D/\epsilon_0$ = 0.8 and 0.9 V/nm, respectively, yielding critical temperature near 100 mK for both states.* ***e****,* ***f****, $R_{xx}$ during forward (dashed red) and backward (solid blue) scans of $H_\perp$ at $n$ = 0.67×$10^{12}$ cm$^{-2}$, $D/\epsilon_0$ = 0.79 V/nm (SC1), and $n$ = 1.0×$10^{12}$ cm$^{-2}$, $D/\epsilon_0$ = 0.9 V/nm (SC2), marked by the yellow triangle and red diamond in* ***b****. In both cases, hysteresis between two zero-resistance states is observed, indicating a ferromagnet-like behavior of both superconducting states.*

Figure 1b shows the longitudinal resistance $R_{xx}$ of device D1 measured at base temperature (10 mK at the mixing chamber) as a function of $n$ and $D$, with a small $\mu_0 H_\perp$ = 50 mT applied to suppress valley fluctuations. Two extended zero-resistance regions are observed and labeled SC1 and SC2. Temperature dependent $R_{xx}$ linecuts as function of $n$ through both regions show that the zero-resistance states are suppressed upon warming and deviating from zero resistance at ~ $100$ mK, confirming true superconducting nature of these states (Figs. 1 c,d and Extended Data Fig. 2). The two superconducting regions appear in the part of the $(n, D)$ phase space that theory calculations suggest an enhanced density of states[19]. The $R_{xx}$ map as a function of $n,\ D$, with two low-resistance regions emerging from the gate-induced flat conduction band, is reminiscent of superconductivity reported in rhombohedral tetra- and pentalayer graphene[4].

A key signature of SC1 and SC2 is their hysteretic response to $H_{\perp}$. Figures 1e,f show $R_{xx}$ during forward and backward sweeps of $H_{\perp}$ at representative $(n, D)$ points inside the SC1 and SC2 regions, marked by the yellow triangle and red diamond in Fig. 1b. In both cases, the system switches between zero resistance states with a non-zero-resistance peak appearing at the coercive field. This behavior is highly unusual for superconductivity and closely resembles the hysteretic superconducting response in CSC in tetra- and pentalayer rhombohedral graphene, where it was associated with time-reversal-symmetry breaking at the orbital level and valley-polarized parent states[4].

## New Superconducting State Under $H_\perp$

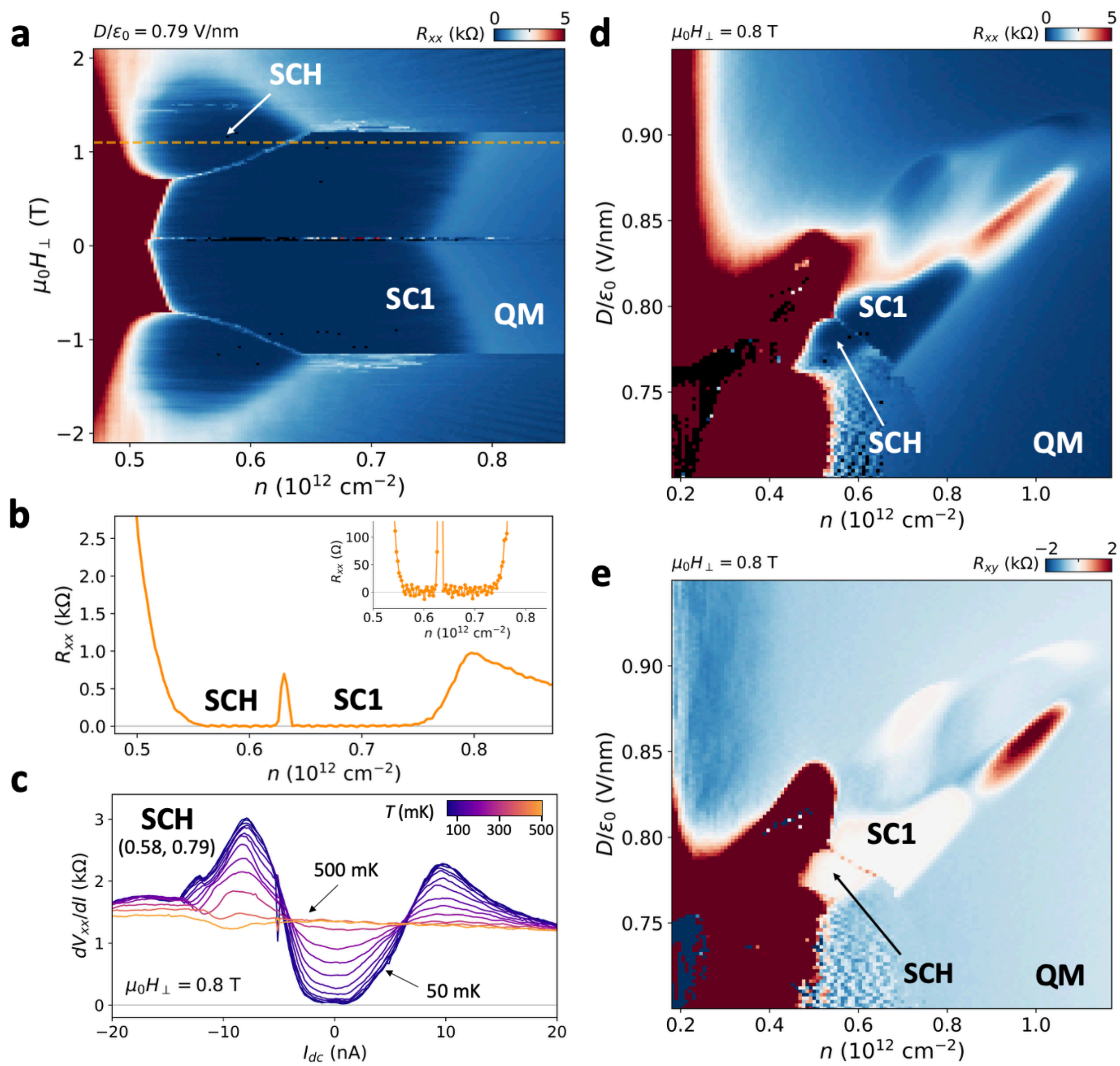

***Fig. 2 | Out-of-plane magnetic field induced superconductivity. a**, $R_{xx}$ as a function of $n$ and $H_\perp$ at $D/\epsilon_0$ = 0.79 V/nm. Above about 0.8 T, a new zero resistance phase emerges, separated from SC1 by a sharp finite resistance phase boundary; we label this new phase SCH. SCH persists up to an upper critical field of approximately 1.6 T, larger than those of SC1 and SC2, as well as the perpendicular critical fields reported for CSC states in thinner rhombohedral graphene.[4] **b**, Linecut along the orange dashed line in **a**, highlighting two distinct zero-resistance states at 1.1 T. Inset: zoomed in view of the same linecut, demonstrating that the resistance is zero within the measurement accuracy of a few ohms. **c**, Temperature and current ($I_{dc}$) dependence of the differential resistance $dV_{xx}/dI$ measured at $n$ = 0.58×$10^{12}$ cm$^{-2}$, $D/\epsilon_0$ = 0.79 V/nm. **d**, **e**, $R_{xx}$ (**d**)*

*and Hall resistance ($R_{xy}$, **e**) maps taken at 0.8 T and base temperature. In **d**, part of the SC1 region at lower $n$ and $D$ is replaced by SCH with a well-defined phase boundary, while SC2 is suppressed at this magnetic field. In **e**, both SC1 and SCH are adjacent to regions exhibiting anomalous Hall signals.*

To examine how superconductivity evolves under $H_{\perp}$, we first map $R_{xx}$ as a function of $n$ and $H_{\perp}$ at fixed displacement field $D/\epsilon_0 = 0.79$ V/nm (Fig. 2a). The most striking observation appears at higher field. Above a threshold of about 0.8 T, a new zero-resistance region emerges at lower $n$ compared to SC1, which we label as SCH. SCH exists over a finite field window and disappears near 1.6 T, corresponding to a much larger perpendicular critical field than the low-field superconducting states SC1 and SC2, as well as previously reported electron-doped superconducting states in thinner rhombohedral graphene[4]. Together with its lower carrier-density range, this enhanced field stability suggests that SCH may lie closer in a strongly coupled superconducting regime. Between approximately 0.8 and 1.2 T, SCH is separated from SC1 by a narrow finite-resistance boundary, as highlighted by the density linecut at $\mu_0 H_{\perp} = 1.1$ T (Fig. 2b). The corresponding negative-field region in Fig. 2a reveals the time-reversal-related counterparts of these superconducting states. Thus, by tuning $n$, $D$, and $H_{\perp}$, R6G provides access to a quartet of superconducting states.

To confirm that SCH is a superconducting state, we measure the differential resistance $dV_{xx}/dI$ as a function of bias current and temperature (Fig. 2c). SCH exhibits a nonlinear current response with a critical-current-like feature and is suppressed upon warming, confirming its superconducting character. We then examine the field-induced state in the $(n, D)$ plane near the threshold field. At $\mu_0 H_{\perp} = 0.8$ T, the $R_{xx}$ map in the $(n, D)$ plane shows that SC2 is suppressed, while SCH occupies part of the lower $n$ side of the phase diagram adjacent to SC1 (Fig. 2d). The corresponding Hall-resistance map $R_{xy}$ (Fig. 2e) shows that both SC1 and SCH have vanishing Hall response within the zero-resistance regions, as expected for superconducting states, while the surrounding metallic regions show non-zero anomalous Hall signals.

## Neighboring QM and Parent States of SC1 and SCH

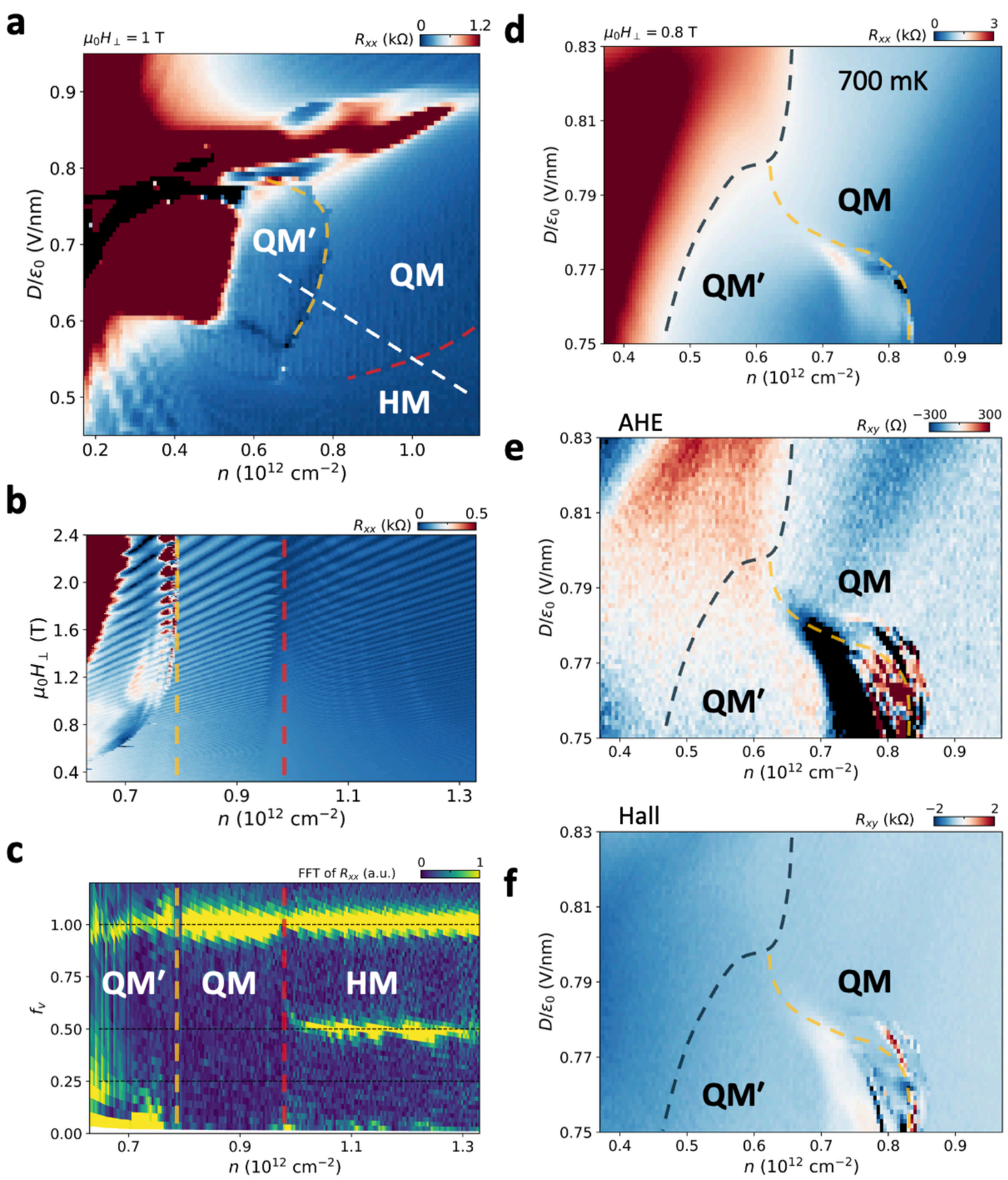

***Fig. 3 | Fermiology of neighboring QM and valley polarization of SCH parent state. a**, Base temperature $R_{xx}$ as a function of $n$ and $D$, measured at 1 T. A new phase labeled by QM′, is induced by $H_\perp$ that is enclosing SCH and separated from QM by a clear boundary, partially indicated by a yellow dashed line. **b**, $R_{xx}$ as a function of $n$ and $H_\perp$ measured along the white dashed line in **a**. **c**, Fourier transform of $R_{xx}(1/H_\perp)$ from **b**, plotted as normalized frequency ($f_\nu$) as a function of $n$. Vertical dashed lines indicate boundaries between HM–QM, and QM–QM′. The oscillation frequency remains unchanged across QM–QM′ boundary, indicating that the phase enclosing SC1 and SCH are both quarter-metal and separated by a sharp boundary. Weak low-frequency spectral weight in the QM′ density range is attributed to non-periodic resistance variations associated with the field induced shift of the QM–QM′ boundary and the nearby high-field insulating state. **d-f**, $R_{xx}$ (**d**), extracted anomalous Hall resistance (**e**) and ordinary Hall*

*resistance (**f**) as a function of n and D measured at 0.8 T and T = 700 mK where SC1 and SCH are fully suppressed. The boundary between QM and QM′ (dashed yellow line) and the adjacent high-resistance state (dashed black line) are extracted from $R_{xx}$ map. In **e**, the anomalous Hall response changes sign across the QM–QM′ boundary, while the ordinary Hall channel in **f** remains unchanged.*

We next examine the fermiology of the metallic states neighboring SC1 and SCH in the $(n, D)$ phase diagram at $\mu_0 H_\perp = 1$ T (Fig. 3a). In addition to the low-field quarter-metal phase QM that is adjacent to SC1, a field-induced metallic phase, which we denote as QM′, emerges and surrounds SCH. QM′ is separated from QM by a clear high-resistance boundary, indicated by the yellow dashed line. Notably, this same boundary connects continuously to the finite-resistance boundary between SC1 and SCH, forming an enclosed region that contains both SCH and the surrounding QM′. The red dashed line marks a separate boundary between the QM regime and a higher $n$ half-metal (HM) phase. The connection between superconductivity and the neighboring metallic states is further supported by their evolution with $H_\perp$. Extended Data Fig. 3d-f shows that, after superconductivity is suppressed at higher fields, quantum oscillations from QM and QM′ appear in the same $(n, D)$ regions that host SC1 and SCH at lower field. The SC1 and SCH contours identified at $\mu_0 H_\perp = 1$ T overlap with the corresponding QM and QM′ metallic regions at $\mu_0 H_\perp = 1.6$ T and 2 T, indicating that SC1 and SCH emerge from these adjacent quarter-metal parent states.

We further identify the fermiology across these neighboring metallic regions using quantum oscillations measured along the white dashed line in Fig. 3a as a function of $n$ and $H_\perp$ (Fig. 3b). The raw and background-subtracted Landau fans show clear quantum oscillations in the same density range where superconductivity appears at lower field (Extended Data Fig. 3a,b). The Fourier transform of $R_{xx}(1/H_\perp)$ shows that the normalized oscillation frequency remains unchanged across the QM–QM′ boundary (Fig. 3c), and the FFT analysis over the superconducting density range shows a dominant quarter-metal frequency with no additional split-frequency branch (Extended Data Fig. 3c). The weak low-frequency spectral weight visible in parts of the QM′ region is likely caused by non-periodic resistance variations from the field-dependent shift of the QM–QM′ boundary and the nearby high-field insulating state, rather than by an additional quantum-oscillation frequency. These results indicate that the metallic regions adjacent to SC1 and SCH both have quarter-metal fermiology with a single non-degenerate Fermi surface, although the precise spin-valley flavor cannot be determined by quantum oscillations alone.

To probe the valley/orbital character of the parent states, we suppress superconductivity by raising the temperature to $T = 700$ mK and measure the Hall signal. In this regime and at $\mu_0 H_\perp = 0.8$ T, SC1 and SCH are replaced by their parent states, while a high-resistance boundary remains visible in $R_{xx}$ at the location corresponding to the QM–QM′ (parent states of SC1 and SCH) boundary (Fig. 3d). We then extract the anomalous Hall signal by subtracting the supposedly linear in $H_\perp$ ordinary Hall signal from the total Hall signal (see Methods). Across this boundary, the anomalous Hall resistance $R_{xy}^{\mathrm{AHE}}$ reverses sign (Fig. 3e), whereas the ordinary Hall resistance remains continuous and electron-like (Fig. 3f). At even higher temperature (4 K) where the quarter-metal parent states disappear, neither the QM–QM′ boundary nor any anomalous Hall signal is observed (Extended Data Fig. 6) as a consistency check for the anomalous and ordinary Hall components separation method. Because the anomalous Hall response in rhombohedral graphene is tied to valley polarization and orbital magnetization[29–32], this sign reversal indicates that the parent states of SC1 and SCH are from opposite valleys and carry opposite orbital magnetization. These results identify SCH not as a simple high-field extension of SC1, but as a superconducting state emerging from a different field induced, isospin polarized QM′ phase. Similar behavior is observed in a second R6G device, supporting the reproducibility of this observation (Extended Data Fig. 7). Unlike the near-zero-field hysteretic switching reported previously, R6G enables two distinct valley-polarized superconducting states to be selected at the same field by tuning $n$ and $D$. Including the time-reversal-related partners selected under negative $H_\perp$, the measured phase space realizes a superconducting quartet associated with all four spin-valley isospin flavors. Whether the parent-state valley/orbital character is directly inherited as superconducting order-parameter chirality remains an open experimental question.

## Kane–Mele SOC and field-driven isospin switching

Motivated by the $H_\perp$ field-induced SCH state and the sign reversal of the anomalous Hall response in the parent states, we consider a minimal phenomenological picture for field-driven isospin switching. The model includes intrinsic Kane–Mele spin–orbit coupling, spin Zeeman coupling and valley-contrasting orbital magnetization. For the rhombohedral graphene conduction band at large $D$, the intrinsic SOC can be written as an effective spin–valley anisotropy,

$$H_{\mathrm{KM}}^{\mathrm{eff}} = -\frac{\lambda_I}{2}\tau_z s_z.$$

Here $\lambda_I$ is the intrinsic strength of the SOC, $\tau_z = \pm 1$ labels the $K/K'$valleys and $s_z = \pm 1$ labels spin[9]. Consistent with first-principle and tight-binding studies of graphene SOC[33,34] , this term produces a small spin–valley splitting[35–38]. It selects a low-energy Kramers pair at zero field, while a magnetic field further splits the spin–valley branches through spin and orbital magnetic moments. In this picture, the observed switching reflects competition between the zero-field spin–valley anisotropy and the field-dependent orbital and spin Zeeman energies. The sign reversal of the anomalous Hall response across the SC1–SCH parent-state boundary favors a switching between opposite valley-polarized branches, rather than a transition involving spin alone.

In an out-of-plane magnetic field, the four spin–valley branches acquire both spin and orbital Zeeman energies. We write the total energy per particle of the quarter-metal (measured with respect to a common ground state energy) as

$$E_{\tau s}(H_\perp) = -\frac{\lambda_I}{2}\tau s - \tau g_{\mathrm{orb}}\mu_B H_\perp + s g_s \mu_B H_\perp,$$

where $g_{\mathrm{orb}}$ and $g_s$ are the valley and spin g-factor respectively. This formula captures the competition between the zero-field spin–valley anisotropy and the field-dependent spin and orbital Zeeman energies. Depending on the relative size of the orbital and spin contributions, two types of level crossing are possible: a spin-switching transition when the orbital contribution dominates, as illustrated in Fig. 4a, or a valley-switching transition when the spin and orbital contributions partially cancel within the low-energy Kramers pair, as illustrated in Fig. 4b.

Our elevated-temperature anomalous Hall measurements reveal a sign reversal across the SC1–SCH parent-state boundary, favoring the scenario (Fig. 4b) in which SC1 and SCH originate from opposite valley-polarized parent states rather than from parent states differing in spin polarization. In this case, the switching happens when $\lambda_I \approx 2\, g_{\mathrm{orb}}\mu_B H_\perp^* \approx 93\, g_{\mathrm{orb}}$ μeV . Microscopic calculations give orbital moments of the order of a few $\mu_B$ in the relevant density range (Extended Data Fig. 8 and Methods). This is also compatible with recent scanning SQUID measurements on rhombohedral multilayer graphene[39,40], which report a small but finite magnetic response in the superconducting regime with a value of approximately $\mu_B$ per carrier. At the same time, the precise shape of the SC1–SCH boundary may not be fully captured by this minimal model. The measured boundary is a superconducting free-energy boundary and may also reflect interaction-enhanced orbital response[41–43], nonlinear field dependence of the effective magnetic moment[44,45], field-induced changes in the Fermi-surface geometry[46–48], and different orbital depairing strengths for superconductors derived from different valleys[49–52]. These additional effects are discussed further in the Methods and will require more systematic future studies. More broadly, the observation that $H_\perp$ selects between distinct quarter-metal parent states provides a framework for understanding how superconductivity, valley polarization and orbital magnetism are coupled in R6G.

**Magnetic-Field Selection and Electrostatic Programming of SC1 and SCH**

The SOC-assisted switching picture provides a natural way to organize the full set of superconducting states observed in R6G. For positive $H_\perp$, the anomalous Hall sign reversal and the minimal spin–valley model suggest that SC1 and SCH originate from quarter-metal parent states with different valley polarization, for example $K' \downarrow$ and $K \downarrow$ branches, respectively. Reversing $H_\perp$ selects the time-reversal-related partners, $K \uparrow$ and $K' \uparrow$ branches. Thus, across the measured $(n, D, H_\perp)$ phase space, superconductivity can be realized from all four spin–valley flavors, establishing a switchable CSC quartet in R6G (Fig. 4c).

The coexistence of SC1 and SCH in R6G enables multi-knob switching between superconducting states using either magnetic field or electrostatic gates. Figure 4d shows a zoomed-in $R_{xx}(n, H_\perp)$ map at $D/\epsilon_0 = 0.79$ V/nm, where SC1 and SCH are separated by a finite-resistance boundary over a broad field range. $R_{xx}$ linecuts taken at $0.9 - 1.2$ T show that tuning $n$ at fixed $H_\perp$ drives a transition between SC1 and SCH, with the transition position shifting systematically as $H_\perp$ is varied (Fig. 4f). More importantly, the same switching can be achieved by electrostatic control at fixed $H_\perp$. In the zoomed-in $(n, D)$ map at 1 T, SC1 and SCH occupy adjacent regions separated by a finite-resistance boundary (Fig. 4e). Linecuts taken over a narrow $D$ range, $D/\epsilon_0 = 0.785 - 0.791$ V/nm, demonstrate that a small change in electric field is sufficient to switch between SC1 and SCH (Fig. 4g).

This electrostatic programmability provides a route to gate-defined domain walls between superconducting states selected from different parent isospin flavors[53]. Unlike magnetic training near zero field, which globally selects between time-reversal-related states, the neighboring SC1 and SCH states can be selected locally under a spatially uniform $H_\perp$ by tuning $n$ or $D$. A split-gate geometry could therefore define spatial boundaries between identical or different superconducting states within a structurally uniform R6G crystal (Fig. 4h). Such domain walls are a natural setting for future transport and phase-sensitive probes of how parent-state isospin flavor is encoded in the superconducting order parameter.

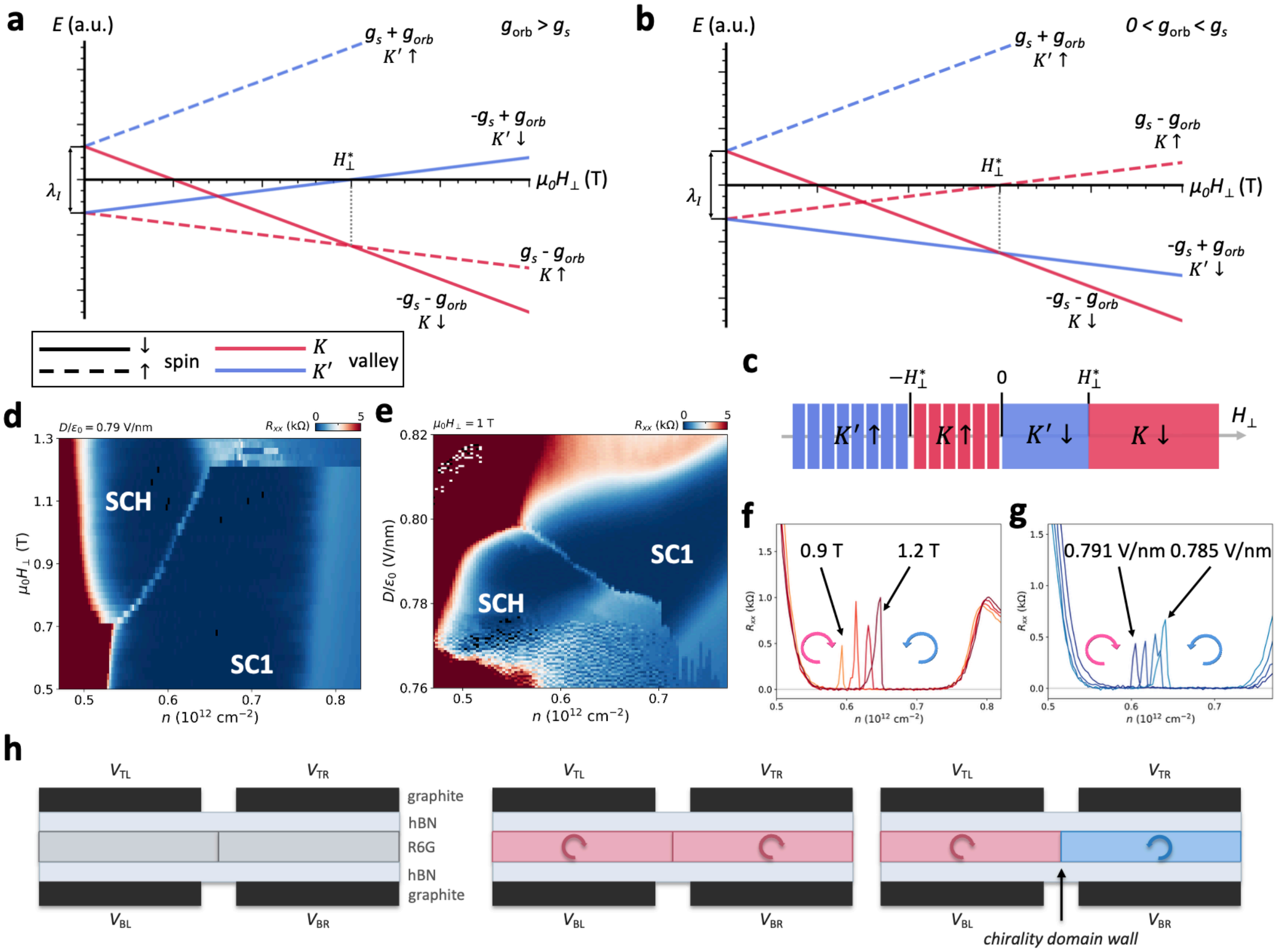

***Fig. 4 | Magnetic-field selection and electrostatic programming of SC1 and SCH. a-b**, Schematic illustration of orbital switching at finite $H_\perp$. At zero field, two degenerate pairs of isospin flavors, $(K\uparrow, K'\downarrow)$ and $(K\downarrow, K'\uparrow)$, are split by $\lambda_I$. The $H_\perp$ shifts their energies according to the combined orbital and spin magnetic moments, and a spin switch occurs when the two lowest-energy states cross, at $2g_s\mu_B H_\perp^* \simeq \lambda_I$ if orbital magnetization $g_{\mathrm{orb}} > g_s$ (**a**) or a valley switch occurs at $2g_{\mathrm{orb}}\mu_B H_\perp^* \simeq \lambda_I$ if $0 < g_{\mathrm{orb}} < g_s$ (**b**). The labels at each curve indicate the slope of the four isospin flavors. Results in Fig 3.**e** favors the scenario in **b**. **c**, Illustration of CSC quartet, chiral superconductors developing from all four isospin polarized quarter–metal parent states can be*

*achieved by only tuning $H_\perp$.* ***d-e***, *$R_{xx}(n, H_\perp)$ at $D/\epsilon_0$ = 0.79 V/nm* ***(d)*** *and $R_{xx}(n, D)$ at 1 T* ***(e)***. ***f-g***, *$R_{xx}(n)$ linecuts at 0.9 T - 1.2 T, in steps of 0.1 T* ***(f)***, *$D/\epsilon_0$ = 0.785 V/nm - 0.791 V/nm, in the increment of 2 mV/nm* ***(g)***, *demonstrating $(n, D, H_\perp)$ programmable switching between SC1 and SCH.* ***h***, *Conceptual split-gate geometry for programmable domain walls between superconducting states in R6G at fixed global $H_\perp$: left, both regions are normal metal; middle, both regions are tuned into the same superconducting state; right, neighboring regions are tuned into different CSC states out of the CSC quartet, such as SC1 and SCH.*

The relation between parent-state valley polarization and superconducting order-parameter chirality remains an important open question. Quantum oscillations establish quarter-metal fermiology for the parent states of SC1 and SCH, and anomalous Hall measurements show that these parent states carry opposite valley polarization. For a quarter metal with an approximately circular Fermi surface, theory generally expects intravalley pairing to favor chiral $p \pm ip$ type superconductivity[5–8], so switching the valley-polarized parent state may switch the favored superconducting chirality. However, the present transport measurements do not directly determine the phase winding of the superconducting gap or establish whether SC1 and SCH have opposite order-parameter chirality. What is experimentally established here is the field- and gate-controlled switching between superconducting states descended from distinct valley polarized quarter-metal parent states. This capability enables gate-defined SC1–SCH junctions within a structurally uniform graphene crystal, providing a platform for phase-sensitive tests of the superconducting order parameter. If future measurements establish that SC1 and SCH are topological CSC with opposite order-parameter chirality, their interface could realize a $p + ip/p - ip$-type domain wall supporting one-dimensional Majorana boundary modes[10,53,54], as illustrated in Fig. 4h.

## Discussion

Our measurements reveal a magnetic-field-induced superconducting state, SCH, in rhombohedral hexalayer graphene and identify a switchable quartet of CSC states associated with the four isospin flavors. At the same sign of $H_\perp$, SCH and SC1 emerge from spin–valley-polarized QM′ and QM parent states, respectively, with the same spin polarization but opposite valley polarization and opposite chirality. By controlling $H_\perp$, $n$ and $D$, we realize highly flexible electromagnetic switching among the four CSC states within the quartet.

The observation of a CSC quartet, together with its electromagnetic controllability, opens new opportunities for exploring the nature of these superconducting states and their possible quantum-device applications. For example, split-gate structures could define stable junctions between SC1, SCH, and the neighboring non-superconducting QM and QM′ states, enabling systematic transport measurements both across and along the domain walls. These gate-defined interfaces would be more stable and controllable than the spontaneously formed, history-dependent magnetic domains imaged by SQUID-on-tip experiments in thinner rhombohedral graphene[39,40].

On one hand, such domain walls could help understanding the unusual phenomenology of this class of unconventional superconductors, including SCH, SC1, and SC2, where Berry-curvature effects and orbital ferromagnetism may have substantial impacts on the superconductivity in rhombohedral graphene. Tunneling through gate-defined metal–insulator-superconductor or Josephson junctions could provide spectroscopic information about the chiral superconducting states and probe valley-dependent superconducting coherence properties at the domain walls[53]. On the other hand, our work could help address the possible chirality and topology of the superconducting order parameter. Majorana boundary modes have been proposed at domain walls between $(p_x + ip_y)$ and $(p_x - ip_y)$ CSC domains[10,11,53], which may inherit opposite valley chirality from the parent states of SC1 and SCH. If this scenario is verified, electromagnetically reconfigurable networks of Majorana boundary modes could be constructed for braiding and quantum-computing applications[10,11,55].

**Data availability**

The data that support the findings of this study are available from the corresponding authors upon request.

**Acknowledgements**

We acknowledge fruitful discussions with Oskar Vafek, Dmitry Chichinadze, Liang Fu, Brian Skinner, and Jihang Zhu. We acknowledge Jackson Butler for help with the transport measurements. P.X. is supported by NSF grant no. DMR-2325147. Z.L. is supported by start-up funding from Florida State University. V.T.P. and C. L. were supported in part by start-up funds from Florida State University and the National High Magnetic Field Laboratory and in part by NSF CAREER grant No. DMR-2543710. The National High Magnetic Field Laboratory is supported by the National Science Foundation through NSF/DMR-2128556 and the State of Florida. Device fabrication was performed at FSU and MIT.nano. K.W. and T.T. acknowledge support from the JSPS KAKENHI (grant nos.20H00354, 21H05233 and 23H02052) and the World Premier International Research Center Initiative, MEXT, Japan.

**Author Contributions**

Z.L., P.X. and L.J. supervised the project. Z.H., S.Y. and G.S. performed the DC magneto-transport measurement. Z.H., S.Y., P.P., E.A., and T.H. fabricated the devices. J.Y., J.S., H.L., R.H., K.X., and J.G. helped with installing and testing the dilution refrigerator. V. T. P. carried out the theoretical simulations supervised by C. H. and C. L. K.W. and T.T. grew hBN crystals. All authors discussed the results and wrote the paper.

**Competing Interests**

The authors declare no competing interests.